\documentclass[%
preprint,
nofootinbib,
 amsmath,amssymb,
 aps,
]{revtex4-1}
\usepackage{graphicx, amsmath, amssymb, amsfonts, pifont, bm, cancel}
\newcommand{\dyad}[1]{\mbox{$\overline{\overline{#1}}$}}

\newenvironment{exer*}
  {\ex}
  {\endex}

\def\be{\begin{equation}}
\def\ee{\end{equation}}
\def\bea{\begin{eqnarray}}
\def\eea{\end{eqnarray}}

\newcommand{\meff}{m_\text{eff}}
\newcommand{\etab}{\eta_{\text{b}}}

\usepackage{xspace} 

\newcommand{\intdA}{\int\!\!\textrm{dA}}
\newcommand{\intdASu}{\int_{S(\textbf{u})}\!\!\textrm{dA}}
\newcommand{\intdAS}{\int_{S}\!\!\textrm{dA}}

\newcommand{\vecb}[1]{\mathbf{#1}}

\newcommand{\Vb}{V_{\text{b}}}
\newcommand{\omegab}{\omega_\text{b}}
\newcommand{\kappam}{\kappa_{\text{m}}}
\newcommand{\omegam}{\omega_{\text{m}}}
\newcommand{\chim}{\chi_{\text{m}}}
\newcommand{\chitot}{\chi_{\text{tot}}}
\newcommand{\Yeff}{Y_{\text{eff}}}
\newcommand{\epsilonr}{\epsilon_{\text{r}}}
\newcommand{\epsilonrij}{\epsilon_{\text{r},ij}}
\newcommand{\epsilonrmu}{\epsilon_{\text{r},\mu}}
\newcommand{\Ohms}{\Omega}
\newcommand{\aout}{a_{\text{out}}}
\newcommand{\alphas}{\alpha_{\text{s}}}
\newcommand{\alphalo}{\alpha_{\text{LO}}}
\newcommand{\omegalo}{\omega_{\text{LO}}}
\newcommand{\Philo}{\Phi_{\text{LO}}}
\newcommand{\phis}{\varphi_{\text{s}}}
\newcommand{\Phis}{\Phi_{\text{s}}}
\newcommand{\Glo}{G_{\text{LO}}}
\newcommand{\kB}{k_{\text{B}}}
\newcommand{\E}{\vecb{E}}
\newcommand{\Eb}{\vecb{E}_{\text{b}}}
\newcommand{\Emu}{\vecb{E}_{\mu}}
\newcommand{\neff}{n_{\text{eff}}}
\newcommand{\deff}{d_{\text{eff}}}
\newcommand{\ngr}{n_{\text{g}}}
\newcommand{\um}{\vecb{u}}
\newcommand{\Qm}{Q_{\text{m}}}
\newcommand{\omegamu}{\omega_{\mu}}
\newcommand{\D}{\vecb{D}}
\newcommand{\keff}{k_{\text{eff}}}

\newcommand{\partialVsq}{\partial_{V^{2}}}
\newcommand{\partialx}{\partial_{x}}
\newcommand{\Deltar}{\Delta_{\text{r}}}
\usepackage{graphicx}
\usepackage{amsmath}
\usepackage{pgfplots}
\usepackage{subfigure}
\usepackage{braket}
\usepackage{xcolor}
\usepackage{tikz}
\RequirePackage{atbegshi}
\usetikzlibrary{positioning}
\usetikzlibrary{arrows}
\usetikzlibrary{decorations.text}

\let\svthefootnote\thefootnote
\newcommand\blankfootnote[1]{%
  \let\thefootnote\relax\footnotetext{#1}%
  \let\thefootnote\svthefootnote%
}
\let\svfootnote\footnote
\renewcommand\footnote[2][?]{%
  \if\relax#1\relax%
    \blankfootnote{#2}%
  \else%
    \if?#1\svfootnote{#2}\else\svfootnote[#1]{#2}\fi%
  \fi
 }

\begin{document}

\preprint{APS/123-QED}

\title{Electrical driving of X-band mechanical waves \\ in a silicon photonic circuit}

\author{Rapha\"{e}l Van Laer, Rishi N. Patel, Timothy P. McKenna, Jeremy D. Witmer and Amir H. Safavi-Naeini}

\affiliation{Department of Applied Physics, and Ginzton Laboratory, Stanford University, Stanford, California 94305, USA$^{\star}$}

\date{\today}

\begin{abstract}
Reducing energy dissipation is a central goal of classical and quantum technologies. Optics achieved great success in bringing down power consumption of long-distance communication links. With the rise of mobile, quantum and cloud technologies, it is essential to extend this success to shorter links. Electro-optic modulators are a crucial contributor of dissipation in such links. Numerous variations on important mechanisms such as free-carrier modulation and the Pockels effect are currently pursued, but there are few investigations of mechanical motion as an electro-optic mechanism in silicon. In this work, we demonstrate electrical driving and optical read-out of a 7.2 GHz mechanical mode of a silicon photonic waveguide. The electrical driving is capacitive and can be implemented in any material system. The measurements show that the mechanically-mediated optical phase modulation is two orders of magnitude more efficient than the background phase modulation in our system. Our demonstration is an important step towards efficient opto-electro-mechanical devices in a scalable photonic platform.
\end{abstract}

\pacs{Valid PACS appear here}
\maketitle
\footnote[]{rvanlaer@stanford.edu}

\newpage

\section{Introduction}

Dissipated energy limits our ability to transmit and process information. Optics plays an essential role in reducing this energy, enabling the long-distance communication links that underpin today's communication networks. Research efforts across the globe envision to transfer this success to shorter links inside data centers, on circuit boards and perhaps on individual chips \cite{Miller2017a}. Energy considerations may even more greatly restrict quantum information processors as many quantum systems require low temperatures to suppress decoherence. Dissipation in a cold environment is severely restricted and limits the transmission rates of microwave-to-optical quantum converters \cite{Pechal2017a}.

Electro-optic modulators are a major source of dissipation in a communication link. Two factors set their energy dissipation: (1) the interaction strength of the electro-optic mechanism and (2) the optical losses of the device in question. Much of the research in photonics attempts to improve these properties, pursuing countless variations on key mechanisms such as free-carrier modulation and the second-order Pockels effect \cite{Reed2014,Miller2017a}. Silicon is a widely used and mature material in photonic integrated circuits \cite{Miller2017a,Selvaraja2009,Rahim2017}. It offers dense integration, low optical loss and promises to leverage existing CMOS infrastructure for fabrication \cite{Soref2006a,Lim2014}. However, silicon is centrosymmetric, so it lacks a strong second-order Pockels effect \cite{Cazzanelli2016}. A variety of hybrid approaches, such as the integration of polymers \cite{Koeber2015,Kieninger2017}, graphene \cite{Sorianello2018} and lithium niobate \cite{Witmer2017a,Weigel2016,Chen2014HybridModulator}, attempt to address silicon's perceived lack of active functionality. In addition, competing thin-film and high-density photonic platforms based on materials such as aluminum nitride \cite{Xiong2012b,Vainsencher2016,Sohn2018,Tadesse2014b}, silicon nitride \cite{Rahim2017,Ji2017,Alexander2018}, diamond \cite{Lee2017,Hausmann2014} and lithium niobate \cite{Poberaj2012a,Zhang2017,Wang2018} are under development.

Here, we explore electrically-excited gigahertz mechanical motion as an effective electro-optic mechanism in a nanoscale photonic waveguide. Previous work shows that mechanical systems couple efficiently to microwave and optical fields in an essentially lossless way \cite{Aspelmeyer2014}. Most efforts focus on either electromechanics or on optomechanics in typically sub-gigahertz mechanical systems \cite{Pitanti2015a,Dieterle2016a,Winger2011d,Aspelmeyer2014,VanAcoleyen2012b}. In this work, we electrically generate and optically detect a gigahertz mechanical mode in silicon. The mechanical mode under study has a frequency ($\approx 7-8$~GHz) in the microwave X-band. It is the same mode that has recently been studied in the context of Brillouin scattering and optomechanics \cite{VanLaer2015,VanLaer2015b,VanLaer2017c}. Our work is also closely related to current electro-optic efforts that harness the third-order Kerr effect in silicon \cite{Timurdogan2017,Cazzanelli2016,Aktsipetrov1999}. In those studies, a constant bias field converts the third-order Kerr effect to an effective second-order Pockels effect. Here, in what is typically called capacitive transduction, a constant bias field converts an oscillating microwave field into an oscillating force at the same frequency. The bias field thus breaks the inversion symmetry of silicon and leads to effective piezoelectricity, enabling direct conversion between microwave photons and phonons. These phonons subsequently generate optical sidebands via silicon's strong photoelasticity \cite{VanLaer2015,VanLaer2015b}.

\begin{figure}[ht]
\includegraphics{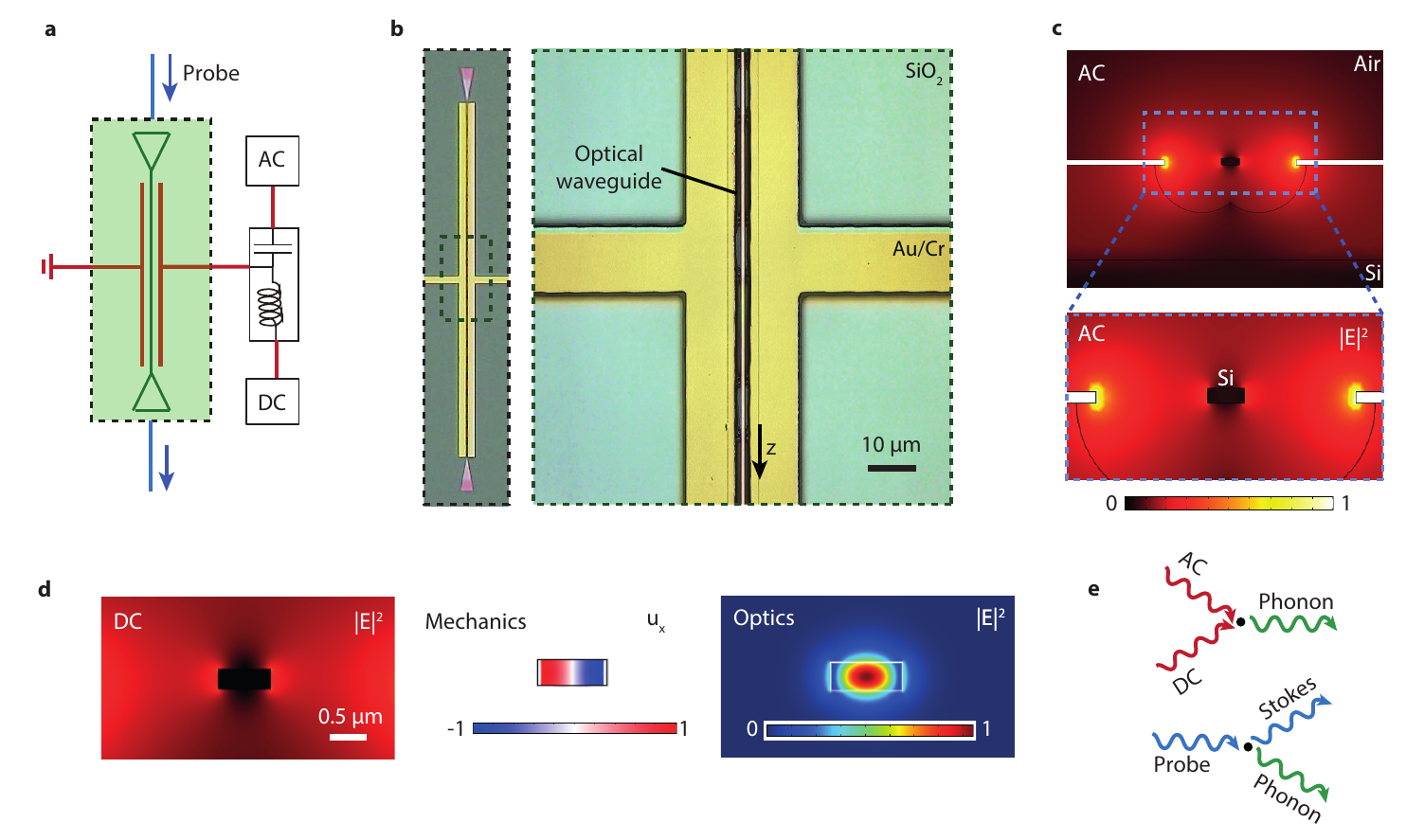}
\caption{\textbf{Electrical driving and optical read-out of gigahertz mechanics in silicon.} \textbf{a}, We electrically drive a gigahertz mechanical mode of a silicon waveguide (green) placed in between the electrodes of a gold capacitor (red). The electrical drive consists of an AC and a DC component. They are combined in an off-chip bias-T. The DC component converts the capacitive force to an effective piezoelectric drive, enabling direct conversion of microwave photons into phonons. The mechanical motion is read out optically via the phase-modulation imprinted on the optical photons (blue) traveling along the waveguide. \textbf{b}, Laser-scanning micrograph of a typical device, showing grating couplers, gold electrodes and a partially suspended silicon waveguide inside the capacitor. The waveguide consists of a series of suspensions held up by silicon dioxide anchors. \textbf{c}, Cross-section of the waveguide in a suspended section, as well as microwave field at 7.2 GHz. \textbf{d}, From left to right: DC electrical field, $\Gamma$-point mechanical mode at 7.2 GHz and optical mode at 193.5 THz with effective index $\neff = 2.49$. \textbf{e}, Principle of the measurement: first, the electrical drive signals generate mechanical motion. Second, the mechanical motion scatters the optical probe into Stokes and anti-Stokes sidebands (only Stokes events shown and measured).}
\label{fig:1}
\end{figure}

\section{Device fabrication}

The device under study is a silicon nanophotonic waveguide in between the electrodes of a gold capacitor. It consists of a series of suspensions to limit mechanical leakage into the thermal oxide \cite{VanLaer2015b}. The fabrication of the device consists of four lithographic steps.

First, we pattern the sub-micron features -- such as the silicon waveguide and grating couplers -- using electron-beam lithography and a Cl$_{2}$/HBr silicon etch into a 220 nm silicon thin-film atop of 3 $\mu$m thermal oxide. The silicon waveguide core is about 580 nm wide. Second, we perform a large-area silicon removal via positive photolithography and another Cl$_{2}$/HBr silicon etch while protecting the photonics structures. This step removes silicon everywhere but in the photonics structures, reducing the risk of dielectric breakdown. Third, we fabricate the gold electrodes via image-reversal photolithography and an electron-beam evaporation of a thin 5 nm chromium adhesion layer and the 165 nm gold electrodes. Fourth, we selectively remove the thermal oxide with positive photolithography and a 6:1 buffered HF etch. In between each step, we perform a thorough 9:1 piranha and 50:1 diluted HF clean. The final piranha/HF clean is shorter to limit etching of the chromium adhesion layer. Finally, we mount the chip on a printed circuit board and wirebond ultrasonically to the traces on the board. We couple optically to the waveguide via $27^{\circ}$ angle-cleaved fibers \cite{Snyder2013} and single-etch focusing metagratings based on \cite{Benedikovic2014}.

The result is a suspended silicon photonic-phononic waveguide in between the electrodes of a gold capacitor atop of thermal oxide (Fig.\ref{fig:1}). Each 3 by 7 mm silicon chip contains eight such devices placed in the same electric circuit in a parallel configuration. We fabricate four such chips simultaneously on a larger 10 by 15 mm silicon piece, which we dice with the final mask already present. We perform the timed release on the individual 3 by 7 mm chips. The devices are typically $L = 546 \, \mu\text{m}$ long, of which about $L_{\text{s}} = 0.71\times L = 388 \, \mu\text{m}$ is suspended. The silicon suspensions and oxide anchors are $17 \, \mu\text{m}$ and $7 \, \mu\text{m}$ long respectively. The gap between the waveguide and the gold electrodes is $1.5 \, \mu\text{m}$ on each side so the gold electrodes induce negligible optical absorption.

\section{Device Physics}
\label{sec:devphys}

We solve for the device's microwave, mechanical and optical fields using finite-element software COMSOL (Fig.\ref{fig:1}c-d). The DC and AC electrical field would be identical were it not for the residual conductivity of the float-zone high-resistivity silicon wafer. Silicon's resistivity $\rho_{\text{Si}} \approx 3 \, \text{k}\Omega \, \text{cm}$ sets an RC-cutoff $\omega_\text{RC}/(2\pi) = 1/(2 \pi \rho_{\text{Si}} \epsilon_{\text{Si}})\approx51\,\text{MHz}$ with $\epsilon_{\text{Si}} = 11.7\epsilon_0$. Oscillating electric fields at frequencies below this cutoff ($\omega \ll \omega_\text{RC}$) do not penetrate the core as they get screened out by the free carriers: silicon acts as a conductor. In contrast, microwave fields at frequencies far above this cutoff ($\omega \gg \omega_\text{RC}$) penetrate the core: silicon acts as a dielectric. Thus the constant bias field $\Eb$ is screened in the silicon, whereas the microwave field $\delta \E$ is merely suppressed by silicon's permittivity.

The key physics of our device consists of two cascaded three-wave mixing processes (Fig.\ref{fig:1}e). On the electromechanical side, the mixing between the bias and microwave field generates mechanical motion. On the optomechanical side, the mechanical motion mixes the optical carrier with its Stokes and anti-Stokes sidebands. Therefore the electro- and optomechanical interactions rates arise from overlap integrals between three fields each (Appendices \ref{sec:electricalexcitation} and \ref{sec:opticalreadout}). Before going to the full model, we consider an approximate qualitative picture that captures much of the relevant physics.

 In the first three-wave mixing process, the bias and microwave fields drive mechanical motion. In particular, the gaps to the electrodes form two capacitors in series whose total capacitance is $C = C_{\text{g}}/2$ with $C_{\text{g}} = \epsilon_{0}A/(g - \delta x)$, $g=1.5\,\mu\text{m}$ and $\delta x$ the mechanical deformation. Since this capacitance, and therefore the electrostatic energy $ Q^{2}/2C$ with $Q=CV$, depends on the mechanical motion $\delta x$, a force
\begin{equation}
\label{eq:capforce}
F = \frac{(\partialx C) V^{2} }{2}= \frac{C V^{2}}{2g}
\end{equation}
of electrical origin is exerted onto the mechanical mode \cite{Jones2013}. As we apply a total voltage $V = \Vb + \delta V$ with $\Vb$ the bias voltage and $\delta V$ the microwave voltage, the force at the frequency of interest scales as $\Vb \delta V$. This sets up a tunable effective piezoelectric drive that generates displacements $\delta x \propto \Vb \delta V$ whose strength is mainly set by $\partialx C$ as well as by the mechanical stiffness and quality factor. The reasoning above captures only the boundary contribution to the electromechanical interaction. We present a derivation for the bulk contribution in Appendix \ref{sec:electricalexcitation}. In this work, the boundary contribution to the electromechanics dominates. Treating the silicon as a dielectric instead of as a conductor in the above has a negligible impact on $\partialx C$ since $\epsilon_{\text{Si}} \gg \epsilon_{0}$.

In the second three-wave mixing process (Fig.\ref{fig:1}e), the electrically-generated mechanical displacement $\delta x$ generates optical phase fluctuations. These fluctuations manifest as Stokes and anti-Stokes sidebands on the optical carrier. This occurs via the modulation of the effective optical refractive index $\neff$. In particular, assuming small phase fluctuations $\delta \phi$ the complex optical amplitude can be expanded as
\begin{equation}
\alphalo e^{i \delta \phi(t)} \approx \alphalo \left(1 + i\delta \phi(t) \right) = \alphalo \left(1 + i k_0 L |\delta \neff| \cos{(\Omega t)} \right)
\end{equation}
with $\delta\phi = k_0 \delta \neff L$, $\alphalo$ the optical carrier amplitude, $k_0$ the vacuum optical wavevector, $L$ the waveguide length and $\Omega$ the mechanical frequency. Thus the electrically-driven phase fluctuations scatter photons into Stokes and anti-Stokes sidebands with an efficiency
\begin{equation}
\label{eq:efficiency}
\eta = \frac{|\delta \phi|^2}{4}
\end{equation}
with $|\delta \phi|=k_0 L |\delta \neff|$ the peak phase fluctuation. We are mainly interested in the phase fluctuations caused by mechanical motion. As in the electromechanical case, the optomechanical interaction has both a boundary and a bulk contribution. In the boundary contribution, the moving material interfaces between silicon and air change the effective optical refractive index $\neff$. In the bulk contribution, the mechanical strain changes silicon's permittivity and therefore the effective index $\neff$ via the photoelastic effect. Contrary to the electromechanical case, the bulk contribution dominates the optomechanical interaction here. We provide self-contained derivations for the optomechanical overlap integrals in Appendix \ref{sec:opticalreadout}.

Besides the phase fluctuations caused by mechanical motion, there are also phase fluctuations generated by the Kerr effect which sets a broadband background beneath the narrowband mechanically-mediated phase modulation. The mechanically-mediated effect is distinguished by its limited bandwidth and strong dependence on waveguide geometry.

\begin{figure}[ht]
\includegraphics{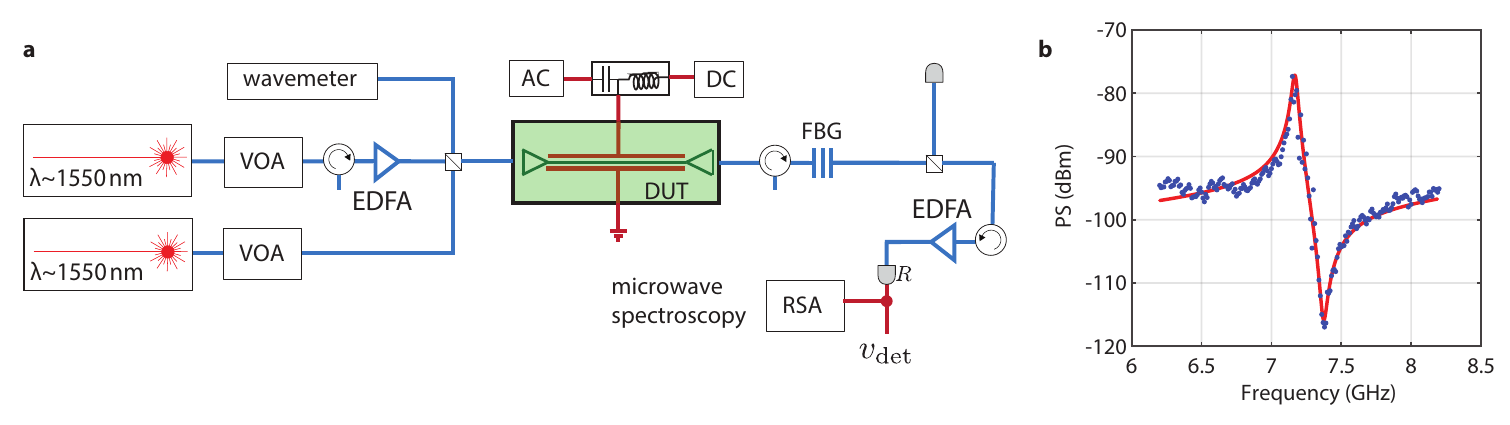}
\caption{\textbf{Measurement setup and typical Fano resonance.} \textbf{a}, We inject 1550 nm laser light into the silicon waveguide to read out the mechanical motion. The mechanical motion generates two sidebands on the optical carrier. The anti-Stokes sideband is rejected by a fiber Bragg grating (FBG), converting the phase- into intensity-modulation. The laser is amplified by erbium-doped fiber amplifiers (EDFAs) to increase the signal-to-noise ratio of the detection. The intensity-modulated signal is sent to an RF spectrum analyzer (RSA) for microwave spectroscopy. The mechanical motion is generated by the electrical drives. Turning off the electrical drives, we calibrate the phase-modulation using a second, heavily attenuated (VOA) laser. We detune this second laser from the main laser by 6.5 GHz using a wavemeter and measure the power spectral density of its beat note with the main laser. This enables calibration of the entire detection chain. The total optical fiber-to-fiber loss is about 20 dB. \textbf{b}, A typical measurement trace: the power spectrum of the photocurrent at the RF frequency as a function of RF frequency. The trace shows a Fano resonance at the mechanical frequency, resulting from the interference between the narrowband mechanical resonance and the broadband Kerr background effect.}
\label{fig:2}
\end{figure}

\section{Measurement Setup}

The goal of our setup is to measure the electrically-induced phase fluctuations $|\delta \phi|$ as a function of applied microwave frequency $\Omega$. We do so in two steps (Fig.\ref{fig:2}a).

First, we turn on the bias and microwave fields and inject 1550 nm laser light into the device. The laser light gets phase modulated and thus has a Stokes and anti-Stokes sideband. We suppress the anti-Stokes sideband by more than 25~dB using a fiber Bragg grating directly after the chip. This partially converts the optical phase to optical intensity fluctuations. Subsequently, we send the carrier and its Stokes sideband to an erbium-doped fiber amplifier and a photodetector. The photodetector generates a photocurrent oscillating at frequency $\Omega$. Finally, we measure the power spectrum of the photocurrent using an electrical spectrum analyzer and determine its peak value. The photodetector and the electrical spectrum analyzer are located in a separate shielded room to minimize microwave crosstalk. This lets us measure signals as low as $150 \, \text{dB}$ below the applied microwave power of about $25 \, \text{dBm}$. We repeat this sequence for a range of microwave frequencies $\Omega$, typically from 6 to 8 GHz with 10 MHz steps. The result is the Fano-shaped curve shown in Fig.\ref{fig:2}b.

Second, we calibrate the measured phase fluctuations $\delta \phi$ \cite{Patel2017c}. To do so, we turn off the electrical drives and inject a second, highly attenuated laser with a known power. We red-detune this laser from the main laser by about 6.5 GHz with both a wavemeter and the electrical spectrum analyzer. Next, we record the power spectral density of the beat note between the two lasers. Using the known power of the second laser and the measured power spectra of both the actual and the calibration photocurrent, we finally determine the absolute magnitude of the phase fluctuations $|\delta \phi|$.

\begin{figure}[ht]
\includegraphics{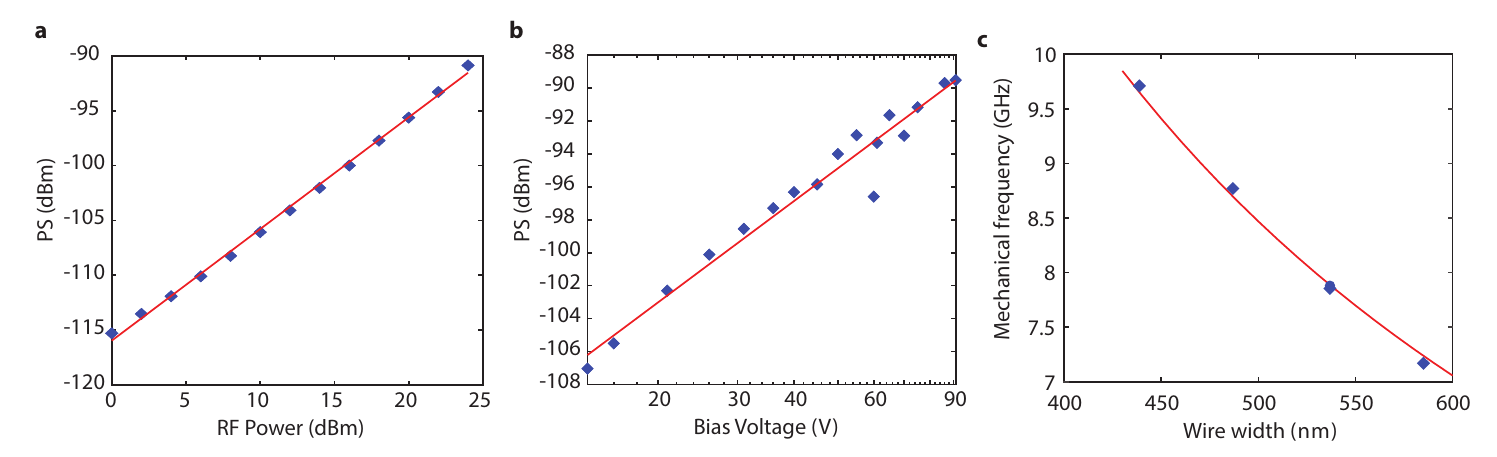}
\caption{\textbf{Scaling of the signal with electrical driving strength and wire width.} \textbf{a}, The detected background signal around 6.5 GHz as a function of microwave drive power (\textbf{a}) and DC bias voltage (\textbf{b}). The red lines are linear fits with slopes of $1.02$ and $2.05$ for \textbf{a} and \textbf{b} respectively, indicating the conversion efficiency scales as $\etab \propto |\delta V \, \Vb|^{2}$ as expected. Thus the bias voltage $\Vb$ converts the capacitive force to an effective piezoelectric drive. \textbf{c}, In addition, we measure the Fano resonance frequencies of a series of devices with varying wire width $w$. The results are in agreement with the expected frequency $\omegam/(2\pi) = v/(2 w)$ of the fundamental Fabry-P\'{e}rot-like mechanical mode (Fig.\ref{fig:1}d).}
\label{fig:3}
\end{figure}

\section{Analysis}

The measurement traces (Fig.\ref{fig:2}b) have a Fano lineshape that fits well to the function
\begin{equation}
\eta = \etab|1 + re^{i\varphi}\mathcal{L}(\Deltar)|^{2}
\end{equation}
with $\etab$ the background conversion efficiency and $\varphi$ an additional phase of the mechanically-mediated phase-modulation with respect to the background Kerr effect. Here we define the Lorentzian
\begin{equation}
\mathcal{L}(\Deltar) =  \frac{1}{-2\Deltar - i}
\end{equation}
with $\Deltar = (\Omega - \omegam)/\kappam$ the relative detuning from the mechanical resonance, $\kappam = \omegam/\Qm$ the mechanical linewidth and $\Qm$ the mechanical quality factor. We give a derivation of this shape in Appendix \ref{sec:opticalreadout}. It stems from the interference between the broadband background phase fluctuations and the narrowband phase fluctuations of mechanical origin. Similar traces have been studied in purely optically-driven cross-phase modulation \cite{VanLaer2015,VanLaer2015b}. The fit yields information on the mechanical quality factor and on the strength of the mechanically-mediated phase fluctuations with respect to the background. We typically obtain
\begin{align}
\Qm &= 167 \\
r &= 9.5 \label{eq:rmeasured}\\
\varphi & = -0.001
\end{align}
The parameter $r$ captures the magnitude of the dimensionless ratio $\delta\phi_{\text{m}}/\delta\phi_{\text{b}}$ between the mechanical and the background phase fluctuations. It gives a voltage-independent measure of how efficient the mechanical mode is with respect to the background Kerr effect at modulating the phase of the optical field. The measurements (Fig.\ref{fig:2}b) show it is up to a factor 10 -- a factor 100 in scattering efficiency $\eta$ -- more efficient at doing so, albeit only in the mechanical bandwidth of roughly $50 \, \text{MHz}$. Further, we measure $\varphi$ close to zero. This implies that the background and mechanically-induced phase fluctuations have the same sign below the mechanical resonance ($\Deltar \ll -1$).

To confirm the physical picture described in section \ref{sec:devphys}, we study the scaling of the measurement traces with three parameters: the microwave power, the bias voltage and the silicon core width (Fig.\ref{fig:3}). First, we find that the power spectrum scales linearly with applied microwave power (Fig.\ref{fig:3}a) and quadratically with applied bias voltage (Fig.\ref{fig:3}b). This is in agreement with the theoretical model. In particular, the power spectrum scales as the conversion efficiency $\eta$. In turn, we have $\eta \propto |\delta\phi|^2 \propto |\delta x|^2 \propto |\delta V \, \Vb|^2 \propto P_{\mu} \Vb^2$ with $P_{\mu} \propto \delta V^{2}$ the microwave power. At small bias voltages, we occasionally measure hysteresis of unknown origin. We take most data at a large bias of $\Vb \approx 80 \, \text{V}$. Second, we find that the Fano resonance frequency is closely predicted by the Fabry-P\'{e}rot frequency $\omegam/(2\pi) = v/(2 w)$ with $v = 8433 \, \text{m/s}$ and $w$ the silicon core width (Fig.\ref{fig:3}c). This agrees with previous all-optical measurements of this mechanical mode \cite{VanLaer2015,VanLaer2015b,VanLaer2017c}.

Our model shows (Appendix \ref{sec:opticalreadout}) that the ratio $r$ is given by
\begin{equation}
\label{eq:ratio}
r =  \frac{\Qm}{\keff} \frac{(\partialx C)(\partialx \neff)}{\partialVsq \neff} \frac{L_{\text{s}}}{L_{\text{b}}}
\end{equation}
with $\keff$ the effective mechanical stiffness and $L_{\text{b}}$ the section of the waveguide that contributes to the background. Here, $\partialx C$ and $\partialx \neff$ capture the electromechanical and optomechanical interaction strengths, while $\partialVsq \neff$ describes the background Kerr interaction strength. Next, we compare the measured to the simulated $r$ as follows. First, we find the background efficiency $\etab$ from the phase-calibrated measurement. Then we compute the measured background interaction strength $L_{\text{b}} \partialVsq \neff$ via an estimate of the applied voltages on the capacitor. This lets us estimate $r$ through our simulated values for the interaction strengths $\partialx C$ and $\partialx \neff$. Thus, from the phase-calibrations we typically measure
\begin{align}
\label{eq:effi}
\etab \approx 3.4 \cdot 10^{-9}
\end{align}
with applied constant bias voltage $\Vb \approx 81 \, \text{V}$ and peak drive voltage $\delta V \approx 5.3 \, \text{V}$. To determine these voltages, we took into account both the microwave cable losses and the electrical response resulting from the wirebonds' inductance, the gold capacitor, the capacitance to the bottom of the chip and the gold resistance (Appendix \ref{sec:electricalexcitation}). We express the background efficiency in terms of the vacuum wavevector $k_0$, the waveguide's $V^2$ background susceptibility $L_{\text{b}}\partialVsq \neff$, the bias voltage $\Vb$ and the peak drive voltage $\delta V$:
\begin{equation}
\etab = \frac{1}{4}\left|k_0 L_{\text{b}}\partialVsq \neff \, \Vb \delta V \right|^{2}
\end{equation}
which along with the measured $\etab$  (equation ~\ref{eq:effi}) yields the estimate
\begin{equation}
\label{eq:freeparam}
L_{\text{b}}\partialVsq \neff \approx \pm 6.7 \cdot 10^{-14} \, \text{m}\cdot\text{V}^{-2}
\end{equation}
Our finite-element simulations predict
\begin{align}
\keff &= 3 \cdot 10^{11} \, \text{N}/\text{m}^2\\
\partialx C &= 0.5 \cdot 10^{-6} \, \text{F}/\text{m}^2 \\
\partialx \neff &= 8.2 \cdot 10^{6} \, /\text{m}
\end{align}
The effective stiffness $\keff$ and optomechanical interaction strength $\partialx \neff$ agree with previous all-optical measurements \cite{VanLaer2015,VanLaer2015b,VanLaer2017c}. Substituting these values in equation \ref{eq:ratio} produces the estimate
\begin{align}
r &\approx \pm 14.0
\end{align}
where the plus-sign holds if the background Kerr effect is positive. The magnitude of the expected $|r| = 14.0$ (equation \ref{eq:rmeasured}) exceeds the measured $r = 9.5$ by $47\%$. We believe that this discrepancy stems largely from the uncertainty in the applied voltages $\Vb$ and $\delta V$ -- which affect the value of $L_{\text{b}}\partialVsq \neff$ in equation \ref{eq:freeparam}. The sign of the expected and measured $r$ are consistent with a positive background Kerr effect ($\partialVsq \neff > 0$).

Finally, we simulate the background Kerr effect from the bulk silicon core and find \begin{equation}
L \partialVsq \neff =  6.3 \cdot 10^{-14} \, \text{m}\cdot\text{V}^{-2}
\end{equation}
if we neglect screening of the constant bias field ($1/\rho_{\text{Si}}\rightarrow 0$). This is close to the experimental value of equation \ref{eq:freeparam} and is in approximate correspondence with measurements of silicon rib waveguides \cite{Timurdogan2017} when taking into account the smaller voltage drop across the waveguide core in our device. To better understand the role of screening, we perform a second set of measurements where the bias field oscillates faster than the RC-cutoff frequency ($\omegab >\omega_{\text{RC}}$). In these sum-frequency driving (SFD) measurements (Appendix \ref{sec:SFD}) we measure much larger Kerr background parameters $L \partialVsq \neff$ and observe no mechanical resonances. Therefore, the SFD measurements suggest at least partial screening of the constant bias field or the presence of additional background mechanisms.

\section{Conclusion}
\label{sec:conclusion}

In conclusion, we demonstrate electrical driving and optical read-out of a 7.2 GHz mechanical mode of a silicon photonic waveguide. The mechanically-driven optical modulation is about two orders of magnitude more efficient than that of the Kerr background. The background is partially screened by the finite silicon conductivity. The screening may be avoided in rib waveguides \cite{Timurdogan2017} or in insulators such as silicon nitride \cite{Grutter2015Si3N4Crystals}. Moving towards propagating instead of localized mechanical modes may improve the electro-optic interaction strength via smaller capacitor gaps that enhance the electromechanical coupling. The absence of piezoelectric materials in this work enables our scheme to be implemented in any material platform, including unreleased silicon-on-insulator \cite{Sarabalis2017Release-freeOptomechanics} and diamond \cite{Lee2017}. Our work shows that an electrical bias field turns silicon into an effective piezoelectric at gigahertz frequencies. These results suggest a route to efficient electro-optic modulation \cite{Miller2017a} and microwave-to-optics quantum conversion \cite{Safavi-Naeini2011e,Andrews2014b,Vainsencher2016,Balram2017} with X-band mechanical waves in scalable photonic circuits. Finally, the introduction of direct and efficient electrical driving of mechanical waves into emerging microwave photonic-phononic systems \cite{Pant,Shin2015,Merklein2016,Sarabalis2017Release-freeOptomechanics,Sohn2018} has the potential to enable a variety of new functions.

\bibliography{Mendeley_Field-induced_piezo.bib}

\vspace{5mm}
\textbf{Acknowledgement.} We acknowledge support by the National Science Foundation (ECCS-1509107, ECCS-1708734), the Stanford Terman Fellowship and the Hellman fellowship, support from ONR QOMAND MURI, as well as start-up funds from the Stanford University school of Humanities and Sciences. R.V.L. acknowledges funding from VOCATIO and from the European Union's Horizon 2020 research and innovation program under Marie Sk\l{}odowska-Curie grant agreement No. 665501 with the research foundation Flanders (FWO). Device fabrication was performed at the Stanford Nano Shared Facilities (SNSF) and the Stanford Nanofabrication Facility (SNF). The SNSF is supported by the National Science Foundation under Grant No. ECCS-1542152. This material is based upon work supported by the National Science Foundation Graduate Research Fellowship under Grant No. DGE-1656518 (R.N.P.). We thank Christopher J. Sarabalis, Patricio Arrangoiz-Arriola, Marek Pechal and Felix M. Mayor for helpful discussions.

\textbf{Contributions.} R.V.L. and A.S.N. conceived of the project. R.V.L. fabricated the devices, performed the simulations, measurements, analysis and wrote the paper. R.N.P. assisted with fabrication, simulations and analysis. T.M.P. and J.D.W. contributed to the design, fabrication and measurement setup. A.S.N. contributed to measurements, simulations, analysis, writing and supervised the project.

\newpage
\appendix

\section{Electrical excitation of mechanical motion}
\label{sec:electricalexcitation}

Here we consider the capacitive excitation of a mechanical oscillator. We send two electrical signals to the electromechanical transducer: a large bias voltage $\Vb(t)$ at frequency $\omegab$ and a small voltage $\delta V(t)$ such that the total voltage is $V = \Vb + \delta V$. Although most of our measurements are made for a constant $\Vb$, we keep the following derivations general for a fluctuating $\Vb(t)$. These voltages generate mechanical motion, which in turn acts back on the capacitance $C$. The capacitor has a stored electrical energy $U = Q^2/(2C)$ with $Q=CV$ the stored charge. The energy $U$ contains the nonlinear capacitance
\begin{equation}
C(V,x) = C_0 + \frac{1}{2}(\partial^2_{V}C) V^2 + (\partial_{x}C) \delta x
\end{equation}
The second term (proportional to $\partial^2_{V}C$) captures the near-instantaneous Kerr effect, while the third term (proportional to $\partial_{x}C$) captures the mechanical motion. The shift in capacitance $\frac{1}{2}(\partial^2_{V}C) V^2$ describes the Kerr effect both in the silicon and in the silicon dioxide. The shift in capacitance $(\partial_{x}C) \delta x$ describes both boundary motion and bulk photoelasticity. Power-conservation dictates that the mechanical force is $F = -\left.\partial_{x}U \right|_{Q}$ with the derivative of the electrical potential energy $U$ evaluated at fixed charge \cite{Jones2013}. The mechanical structure thus experiences a force $F = -\left.\partial_{x}U \right|_{Q} = \frac{Q^2}{2 C^2} \partial_{x}C = (\partial_{x}C) V^2/2$ so its dynamics is given by
\begin{equation}
\label{eq:mech}
\delta \ddot{x}(t) + \kappam \delta \dot{x}(t) + \omegam^2 \delta x(t) = \frac{F(t)}{\meff}
\end{equation}
with $\kappam$ the mechanical linewidth, $\omegam$ the mechanical resonance frequency and $\meff$ mechanical modal mass. From here on, we take $V^2 = \Vb^2 + 2\Vb \delta V + \delta V^2 \approx 2 \Vb \delta V$ -- assuming that the other terms are either negligible or mismatched from the mechanical resonance frequency $\omegam$. Thus we have
\begin{equation}
\label{eq:linearizedC}
C(V(t),x(t)) = C_0 + (\partial^2_{V}C) \Vb(t) \delta V(t) + (\partial_{x}C) \delta x(t)
\end{equation}
Fourier-transforming \ref{eq:mech} we get
\begin{equation}
\left(\omegam^2 - \omega^2 - i \omega \kappam\right)\delta x(\omega) = \frac{(\partial_{x}C)}{2\pi\meff} \Vb(\omega) \star \delta V(\omega)
\end{equation}
Next, we take $\Vb(t) = |\Vb|\cos{(\omegab t)}$ so the convolution $\Vb(\omega) \star \delta V(\omega) = \pi |\Vb|\left(\delta V (\omega - \omegab) + \delta V\left(\omega + \omegab\right)\right)$. The term in $\delta V (\omega - \omegab)$ corresponds to sum-frequency driving (SFD) of the mechanical oscillator, while the term in $\delta V (\omega + \omegab)$ corresponds to difference-frequency driving. We focus here on SFD, assuming negligible Fourier-components $\delta V (\omega + \omegab) = 0$ in a range of frequencies $\omega \approx \omegam \pm \kappam$. We presume $\delta V(\omega - \omegab)$ to be strong in an interval $\omega \approx \omegam \pm \kappam$. Thus we get
\begin{equation}
\delta x(\omega) = \frac{1}{2} \chim(\omega)(\partial_{x}C) |\Vb|\delta V(\omega - \omegab)
\end{equation}
with the mechanical susceptibility
\begin{align}
\chim(\omega) = \frac{1}{\meff\left(\omegam^2 - \omega^2 - i \omega \kappam\right)}
\end{align}
with $\keff = \meff\omegam^{2}$. The Fourier-transform of the capacitance \ref{eq:linearizedC} is
\begin{align}
C(\omega) &= C_0 \delta(\omega) + (2\pi)^{-1}(\partial^2_{V}C)\Vb(\omega) \star \delta V(\omega) + (\partial_{x}C) \delta x(\omega) \\
&= C_0 \delta(\omega) + \frac{1}{2}\left((\partial^2_{V}C) + \chim(\omega)(\partial_{x}C)^2\right)|\Vb|\delta V(\omega - \omegab) \\
&= C_0 \delta(\omega) + \chitot(\omega)\delta V(\omega - \omegab)
\end{align}
with the total susceptibility of the capacitance to voltage defined as
\begin{equation}
\chitot(\omega) = \frac{1}{2} |\Vb|\left((\partial^2_{V}C) + \chi_{\text{m}}(\omega)(\partial_{x}C)^2\right)
\end{equation}

\subsection{Effective mechanical impedance}

The current $\delta I$ flowing through the capacitor is
\begin{align}
\delta I(t) &= \frac{\text{d}}{\text{d}t}(C(t) V(t)) \\
&= C \dot{V} + \dot{C}V \\
&\approx C_0 \dot{V} + \dot{C}\Vb
\end{align}
The capacitance has the strongest Fourier-components around $\omega \approx \omegam$. The $\dot{C}\Vb$ term converts these components back to $\omega\approx\omegam - \omegab$:
\begin{align}
\delta I(\omega) &= -i\omega \left( C_0 \delta V(\omega) + (2\pi)^{-1} C(\omega)\star \Vb(\omega)\right) \\
&= -i\omega\left(C_0 + \frac{|\Vb|}{2} \chitot(\omega + \omegab) \right) \delta V(\omega) \\
&= \Yeff(\omega) \delta V(\omega)
\end{align}
where we focused on the terms in $\delta V$ and defined the effective admittance
\begin{align}
\Yeff(\omega) &= -i\omega\left(C_0 + \frac{|\Vb|}{2} \chitot(\omega + \omegab) \right) \\
&= -i\omega \left(C_0 + \frac{|\Vb|^2}{4} \left((\partial^2_{V}C) + \chi_{\text{m}}(\omega+\omegab)(\partial_{x}C)^2\right) \right)
\label{eq:Yeff}
\end{align}
which includes the dynamical back-action onto the electrical circuit. This result is also valid for a constant bias voltage $\Vb$ with $\omegab=0$. For our current device the back-action terms in equation \ref{eq:Yeff} are negligible such that $\Yeff(\omega) = - i\omega C_{0}$ to a good approximation. Next, we develop expressions for the electromechanical interaction strength $\partialx C$.

\subsection{Electromechanical overlap integrals}

We relate the electromechanical coupling strength $\partial_{x} C$ to the microwave field $\Emu$ and mechanical field $\textbf{u}$ through the surface integral
\begin{equation}
C = \frac{\epsilon_0 L}{|V|^{2}} \intdASu \, \Emu^{\star} \cdot \dyad{\epsilonrmu}(\Emu,\textbf{u}) \cdot \Emu
\end{equation}
over a cross-section $S$ with $L$ the electrode length, $\epsilonrmu$ the relative microwave permittivity and $V$ the voltage that generates the microwave field $\Emu$. The Kerr effect perturbs $\epsilonr$ inside the bulk silicon and silicon dioxide. Mechanical motion perturbs $C$ through $S(\textbf{u})$ via shifts in the boundaries between the constituent materials. It also generates a strain in the bulk silicon which shifts $\epsilonr$ through the photoelasticity. The latter is converted to an induced piezoelectricity through the bias voltage $\Vb$. Note that here we are interested in changes in the microwave permittivity and energy, not in the optical permittivity or energy.

First, for the bulk contribution to $\partialx C$ we have
\begin{align}
\left.\dyad{\delta \epsilonrmu} \right|_{\text{p.e.}} &= - \epsilonr^{2} \, \dyad{p}_{\mu} \cdot \dyad{S} \notag \\
& = - \epsilonr^{2} \, \dyad{p}_{\mu} \cdot \dyad{s} \, \, \delta x 
\end{align}
with the microwave-frequency photoelastic tensor $\dyad{p}_{\mu}$, the normalized strain
\begin{equation}
s_{kl} = \frac{1}{2}\left(\partial_{k}q_{l} + \partial_{l}q_{k} \right)
\end{equation}
and the normalized displacement field $\vecb{q} = \um/\text{max}(\um)$. From here on, we normalize the $\partialx C$ per unit waveguide length $L$. Thus we have
\begin{align}
\label{eq:petoem}
\left.\partialx C\right|_{\text{p.e.}} = -\frac{\epsilonr^{2} \epsilon_0}{|V|^{2}} \intdAS \, \Emu^{\star} \cdot \left(\dyad{p}_{\mu} \cdot \dyad{s} \right) \cdot \Emu
\end{align}

Second, for the boundary contribution to $\partialx C$ we have
\begin{equation}
\intdASu \, \Emu^{\star} \cdot \dyad{\delta \epsilonrmu} \cdot \Emu = \epsilon_0^{-1} \delta x \int_{\mathcal{C}} \text{d}l \, q_{\text{n}}[\Omega] \, \left(\Delta \epsilon_{\mu} |\E_{\mu||}|^{2} - \Delta \epsilon_{\mu}^{-1} |\D_{\mu\perp}|^{2} \right)
\end{equation}
with $\mathcal{C}$ a curve capturing the interfaces, $q_{\text{n}}$ the component of the normalized displacement field $\vecb{q}$ normal to the interface and pointing towards the medium with permittivity $\epsilon_{\text{o}}$, $\Delta \epsilon = \epsilon_{\text{i}} - \epsilon_{\text{o}}$, $\Delta \epsilon^{-1} = \epsilon^{-1}_{\text{i}} - \epsilon^{-1}_{\text{o}}$ the changes in permittivity at the interfaces, $\E_{||}$ the electrical field parallel to the interface and $\D_{\perp} = \epsilon \E_{\perp}$ with $\E_{\perp}$ the electrical field perpendicular to the interface. Therefore,
\begin{equation}
\label{eq:mbtoem}
\left.\partialx C\right|_{\text{m.b.}} = \frac{1}{|V|^{2}} \int_{\mathcal{C}} \text{d}l \, q_{\text{n}}[\Omega] \, \left(\Delta \epsilon_{\mu} |\E_{\mu||}|^{2} - \Delta \epsilon_{\mu}^{-1} |\D_{\mu\perp}|^{2} \right)
\end{equation}

The above equations \ref{eq:petoem} and \ref{eq:mbtoem} concern the intra-modal coupling between one and the same microwave field $\Emu$ through the mechanical degree of freedom $\delta x$. In our case, this microwave field $\Emu = \Eb + \delta\E$ consists of a bias field $\Eb$ and a fluctuation $\delta\E$ generated by the voltages $\Vb$ and $\delta V$. We are interested in the generation of mechanical waves through the sum-frequency beat note between $\Eb$ and $\delta \E$. The bulk overlap integral in fact becomes
\begin{align}
\left.\partialx C\right|_{\text{p.e.}} = -\frac{\epsilonr^{2} \epsilon_0}{|\Vb \delta V|} \intdAS \, \Eb \cdot \left(\dyad{p}_{\mu} \cdot \dyad{s} \right) \cdot \delta \E
\label{eq:es}
\end{align}
with
\begin{align}
&\sum_{ijkl}p_{\mu ijkl}E_{\text{b},i}\delta E_{j} s_{kl} = p_{\mu 11}E_{\text{b},x} \delta E_{x} s_{xx} \\
& p_{\mu11} E_{\text{b},y} \delta E_{y} s_{yy} + p_{\mu12} E_{\text{b},y} \delta E_{y} s_{xx} \notag \\
& + p_{\mu 12} E_{\text{b},x} \delta E_{x} s_{yy} + 2p_{\mu 44}\Re{(E_{\text{b},x} \delta E_{y})} s_{xy} \notag
\end{align}
where we used $s_{xz} = s_{yz} = s_{zz} = 0$ for our $\Gamma$-point mechanical mode. Similarly, the moving boundary integral in fact becomes
\begin{align}
\label{eq:mb}
\left.\partialx C\right|_{\text{m.b.}} = \frac{1}{|\Vb \delta V|} \int_{\mathcal{C}} \text{d}l \, q_{\text{n}}[\Omega] \, \left(\Delta \epsilon_{\mu} \E_{\text{b},||} \cdot \delta \E_{||} - \right. \\
\left.\Delta \epsilon_{\mu}^{-1} \D_{\text{b},\perp} \cdot \delta \D_{\perp} \right) \notag
\end{align}
In our device, the horizontal displacement dominates the overlap integral. Assuming the bias field vanishes in the bulk silicon, we have
\begin{align}
\label{eq:mbourcase}
\left.\partialx C\right|_{\text{m.b.}} &= -\frac{1}{|\Vb \delta V|} \int_{\mathcal{C}} \text{d}l \, q_{x}[\Omega] \, \left(\Delta \epsilon_{\mu}^{-1} \D_{\text{b},x} \cdot \delta \D_{x} \right) \\
&= \frac{1}{|\Vb \delta V|} \int_{\mathcal{C}} \text{d}l \, q_{x}[\Omega] \, \epsilon_{0} \E_{\text{b},x} \cdot \delta \E_{x}
\end{align}
to a good approximation. The total electromechanical coupling is $\partialx C = \left.\partialx C \right|_{\text{p.e.}} + \left.\partialx C \right|_{\text{m.b.}}$.

\subsection{Bulk vs. boundary contributions to the electromechanical interaction}

As shown in equations \ref{eq:es} and \ref{eq:mb}, the bias field $\Eb$ converts both bulk and boundary capacitive forces into tunable effective piezoelectric forces. In this section, we roughly estimate the maximum strength of this induced piezoelectricity in bulk silicon. We neglect anisotropy and look for order-of-magnitude estimates.

In a piezoelectric material, an oscillating electrical field $\delta E$ linearly transduces an oscillating strain $S$ given by
\begin{equation}
S = d \cdot \delta E
\end{equation}
with $d$ the piezoelectric coefficient. Similarly, in an electrostrictive material (i.e. all materials) an oscillating product $E_{\text{b}}\delta E$ of two electrical fields causes an oscillating stress $T$ given by
\begin{equation}
T = -\epsilon_{0} \epsilonr^{2} p_{\mu} E_{\text{b}}\delta E
\end{equation}
with $n$ the refractive index and $p_{\mu}$ the photoelastic coefficient. This stress $T$ is accompanied by a strain $S = kT$ with $k$ the stiffness coefficient. Therefore, any material exposed to a bias field $E_{\text{b}}$ contains a linear coupling between electrical field fluctuations $\delta E$ and strain given by
\begin{equation}
S = \deff \cdot \delta E
\end{equation}
with the induced piezoelectric coefficient
\begin{equation}
\deff = -\frac{\epsilon_{0} \epsilonr^{2} p_{\mu} E_{\text{b}}}{k}
\end{equation}
Thus we can compare $\deff$ directly to $d$ to see whether the electrostrictive material may outperform the piezoelectric material. There is always a bias field for which $\deff > d$, but the required bias field may cause dielectric breakdown or be impractically large. Next, we insert values for silicon. The microwave photoelastic coefficient $p_{\mu}$ of silicon has not been measured to the best of our knowledge. Density-functional theory calculations \cite{Hounsome2006} and extrapolations from measurements at optical frequencies \cite{Biegelsen1975} expect the photoelasticity to be similar or slightly smaller at microwave than at optical frequencies. Using $\epsilonr = 11.7$, $p_{\mu}\approx-0.09$ and $k \approx 130 \, \text{GPa}$ we find
\begin{equation}
\deff \approx 8.4 \cdot 10^{-4} (E_{\text{b}}\,[\text{V}/\mu\text{m}]) \, \text{pm/V}
\end{equation}
Therefore the bias field must be of order $10 \, \text{kV}/\mu\text{m}$ to match the largest piezoelectric coefficients of a common piezoelectric material such as lithium niobate with $d \approx 15 \, \text{pm/V}$ \cite{Weis1985a}, whereas it must be only of order $1 \, \text{kV}/\mu\text{m}$ to match those of aluminum nitride with $d \approx 1 \, \text{pm}/\text{V}$ \cite{Dubois1999}. Although these fields are large, whether they are feasible depends on device details \cite{McKay1954}.

In this work, with applied bias voltages of $\Vb = 80$ we anticipate fields of $E_{\text{b}} = 3.3 \, \text{V}/\mu\text{m}$ inside the core assuming vanishing silicon conductivity. Then we have $\deff \approx 3 \, \text{fm}/\text{V}$. With a fluctuating voltage of $\delta V = 2 \, \text{V}$ and field of $\delta E \approx 0.1 \, \text{V}/\mu\text{m}$ this yields strain of $S \approx 3 \cdot 10^{-10}$ and displacement of the order $\delta x \approx 0.1 \, \text{fm}$. Using equation \ref{eq:capforce} this corresponds to $\left.\partialx C\right|_{\text{p.e.}} \approx 0.1 \cdot 10^{-6} \, \text{F}/\text{m}^2$ -- in agreement with an estimate based on equation \ref{eq:es}. For the boundary contribution we compute $\left.\partialx C\right|_{\text{m.b.}} = 0.5 \cdot 10^{-6} \, \text{F}/\text{m}^2$ with the finite-element method. However, as soon as we take into account the residual conductivity of silicon, we have $E_{\text{b}} = 0$ inside the core and thus $\partialx C = \left.\partialx C \right|_{\text{m.b.}}$ as $\left.\partialx C\right|_{\text{p.e.}} = 0$. We conclude that in this work the boundary electromechanical interaction dominates and $\partialx C = \left.\partialx C \right|_{\text{m.b.}} = 0.5 \cdot 10^{-6} \, \text{F}/\text{m}^2$. We compute $\left.\partialx C \right|_{\text{m.b.}}$ both via overlap integrals given in equation \ref{eq:mbourcase} and via direct perturbation of the structure in a finite-element model and find agreement to within $10\%$.

\begin{figure}[ht]
\includegraphics{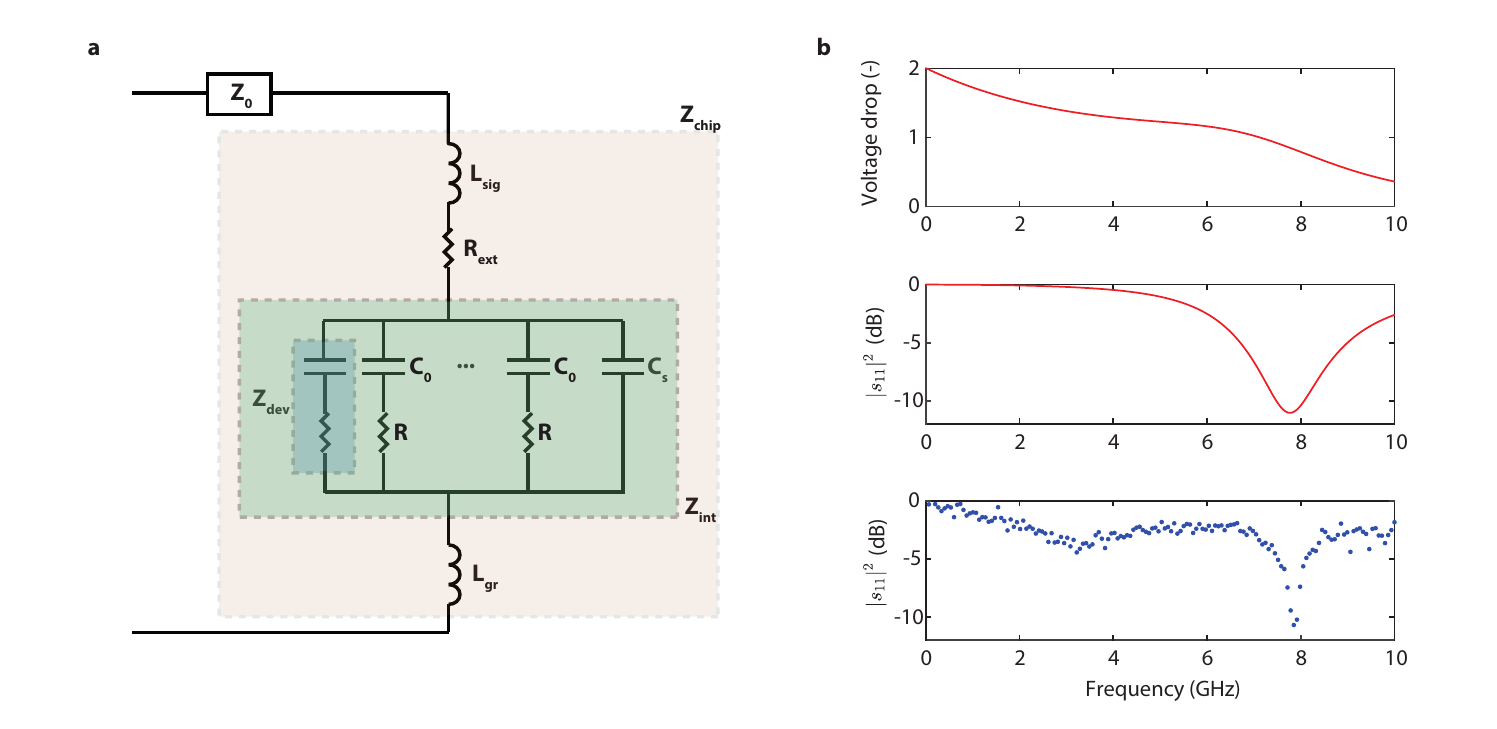}
\caption{\textbf{Electrical response of the chip.} \textbf{a}, Lumped-element equivalent electrical circuit of the chip. We include the wirebond inductance, the gold capacitors $C_{0}$, the capacitance to the bottom of the chip $C_{\text{s}}$ and the gold resistance $R$ and $R_{\text{ext}}$. \textbf{b}, From top to bottom: the simulated microwave voltage drop across each device $Z_{\text{dev}}$, the simulated microwave reflection $|s_{11}|^{2}$ and the measured microwave reflection $|s_{11}|^2$ as a function of frequency. The measured reflection exhibits a $\approx 10 \, \text{dB}$ dip at the LC-resonance of the circuit shown in \textbf{a}. We take into account the measured cable losses in the top figure.}
\label{fig:electricalresponse}
\end{figure}

\subsection{External electrical circuit}

Next, we consider the larger electrical circuit around our device (Fig.\ref{fig:electricalresponse}). We connect a coaxial cable with characteristic impedance $Z_0 = 50 \, \Ohms$ to our chip via a matched coplanar waveguide on a printed circuit board and millimeter-scale wirebonds. On the chip, we connect $N=8$ similar devices in parallel. The chip is mounted on a printed circuit board such that its bottom surface is also at electrical ground. We model each device with an impedance $Z_{\text{dev},k} = Z_{\text{eff},k} + R_{k}$ with $Z_{\text{eff},k} = Y_{\text{eff},k}^{-1}$ given by equation \ref{eq:Yeff} in series with a resistance $R_{k}$. Also taking into account the capacitance $C_{\text{s}}$ to the bottom of the chip, the $N$ devices in parallel have an impedance $Z_{\text{int}}$ set by
\begin{equation}
Z_{\text{int}}^{-1} = \sum_{k=1}^{N} Z^{-1}_{\text{dev},k} - i\omega C_{\text{s}} \approx N Z^{-1}_{\text{dev}} - i\omega C_{\text{s}}
\end{equation}
Here we use $Z_{\text{dev},k} \approx Z_{\text{dev}} \, \forall k$, neglecting small differences in the individual device parameters for simplicity. The devices are connected to the printed circuit board via wirebonds and on-chip electrode pads. We model the impedances to the signal and ground traces $Z_{\text{sig}}$ and $Z_{\text{ground}}$ associated with these connections as a series inductor and resistor $Z_{\text{ext}} = -i\omega L_{\text{ext}} + R_{\text{ext}}$ with $L_{\text{ext}} = L_{\text{sig}} + L_{\text{gr}}$. The total impedance of the chip and its wirebonds is $Z_{\text{chip}} = Z_{\text{ext}} + Z_{\text{int}}$. We model the voltage drop as
\begin{equation}
\frac{\delta V_{\text{dev}}}{\delta V^{+}_{\text{ext}}} = \frac{2Z_{\text{int}}}{Z_{\text{chip}}+Z_{0}} \cdot 10^{-\frac{\omega}{2 \omega_{\text{LC}}}}
\end{equation}
with $\delta V^{+}_{\text{ext}}$ the incident voltage, $\omega_{\text{LC}}/(2\pi) = 7.9 \, \text{GHz}$ the LC resonance frequency (Fig.\ref{fig:electricalresponse}b) and where the last factor takes the microwave cable losses roughly into account. In addition, the microwave reflection is given by
\begin{equation}
s_{11} = \frac{Z_{\text{chip}} - Z_{0}}{Z_{\text{chip}} + Z_{0}}
\end{equation}
The simulated $|s_{11}|^{2}$ is in approximate agreement with the measured $|s_{11}|^{2}$ (Fig.\ref{fig:electricalresponse}b). In our simulations, we use the estimates
\begin{align}
L_{\text{sig}} &= L_{\text{gr}} = L_{\text{ext}}/2 \\
L_{\text{ext}} &= 2.4 \, \text{nH} \\
C_{0} &= 16.4 \, \text{fF} \\
C_{\text{s}} &= 46.2 \, \text{fF} \\
\rho_{\text{Au}} &= 8 \cdot 10^{-8} \, \Omega \text{m} \\
R_{\text{dev}} &= 32.2 \, \Omega \\
R_{\text{ext}} &= 24.1 \, \Omega
\end{align}
with $\rho_{\text{Au}}$ the resistivity of the evaporated Au/Cr thin-film.

\section{Optical read-out of mechanical motion}
\label{sec:opticalreadout}

\subsection{Heterodyne measurement of optical phase fluctuations}

The electrical signals generate changes in the waveguide's optical permittivity via the Kerr effect and mechanical motion. These changes in permittivity phase-modulate the optical field, generating two sidebands. We filter out one of the sidebands using a fiber Bragg grating (FBG) with a suppression exceeding $25 \, \text{dB}$ and a flank width of $2.5 \, \text{GHz}$. The resulting complex signal is
\begin{equation}
\label{eq:opticsfluctuation}
\aout = \alphalo + \delta \alpha + \xi
\end{equation}
with $\alphalo$ the local oscillator amplitude, $\delta \alpha = \alphas e^{-i \Omega t}$ the electrically induced fluctuations, $\Omega$ the modulation frequency, $\alphas$ the signal and $\xi$ the shot noise. We choose $\alphalo$ to be real and treat $\xi$ as a quantum fluctuation. This field generates a photocurrent $I = R (\hbar \omegalo)  \aout^{\dagger}\aout$ with $R$ the photodetector's responsivity and $\omegalo$ the carrier frequency. Thus we have
\begin{equation}
I = R (\hbar \omegalo) \left\{\Philo + \alphalo \left( \delta \alpha+ \delta \alpha^{\star} + \xi + \xi^{\dagger}\right)\right\}
\end{equation}
with $\Philo = |\alphalo|^2$ the carrier photon flux. The autocorrelation of the photocurrent is
\begin{equation}
\label{eq:currentcorrelator}
\begin{split}
&\langle I(\tau) I(0)\rangle = R^2 (\hbar \omegalo)^{2}\Big\{ \langle \Philo(\tau) \Philo(0) \big. \\ 
& + \Big. 4\Philo \left( \langle  \Re{\delta\alpha(\tau)} \Re{\delta \alpha(0)} \rangle + \frac{\delta(\tau)}{4}\right)\Big\}
\end{split}
\end{equation}
where we used $\langle \xi(\tau)\xi^{\dagger}(0) \rangle = \delta(\tau)$ for the shot-noise and other cross-terms vanish as they annihilate the vacuum. Next, we set $\delta \alpha = |\alphas|e^{-i\Omega t - \phis}$ such that
\begin{align}
 \langle  \Re{\delta\alpha(\tau)} \Re{\delta \alpha(0)} \rangle &= |\alphas|^2 \langle\cos{(\Omega \tau + \phis) \cos{(\phis)}}\rangle \\
 &= \frac{|\alphas|^2}{2}\left(\langle \cos{(\Omega \tau)}\rangle + \langle\cos{(\Omega \tau + 2\phis)} \rangle \right) \\
 &= \frac{\Phis}{2} \cos{(\Omega \tau)}
\end{align}
where we defined the signal photon flux $\Phis = |\alphas|^2$ and the term containing $2\phis$ averages out as there is no absolute timing reference. Dropping the $\Philo^2$ term in \ref{eq:currentcorrelator}, the autocorrelation of the photocurrent is
\begin{equation}
\begin{split}
\langle I(\tau) I(0)\rangle = \Glo \left(2\Phis  \cos{(\Omega \tau)} + \delta(\tau)\right)
\end{split}
\end{equation}
with $\Glo = R^2 (\hbar \omegalo)^{2} \Philo$ the measurement gain. Therefore, the power spectral density of the photocurrent is
\begin{align}
S_{II}(\omega) &= \int^{\infty}_{-\infty} \text{d}\tau \, e^{i\omega \tau} \langle I(\tau) I(0)\rangle \\
&= 2\pi\Glo \left( \Phis \left(\delta(\omega - \Omega) + \delta(\omega + \Omega)\right) + 1\right)
\end{align}
The electrical spectrum analyzer measures $Z S_{II}(\omega)$ over a resistor $Z$. Integrating the spectral density over a bandwidth $\Delta \omega$ we obtain
\begin{equation}
\label{eq:noisepower}
\int^{\Omega + \Delta \omega/2}_{\Omega - \Delta \omega/2} \frac{\text{d}\omega}{2\pi} \, Z S_{II}(\omega) = Z \Glo \left( \Phis + \Delta \omega \right) + 2 \kB T \Delta \omega
\end{equation}
where we added the Johnson-Nyquist noise associated with $Z$ in the last term. Therefore, the signal-to-noise ratio is
\begin{equation}
\text{SNR}=\frac{\Phis}{\Delta \omega}
\end{equation}
assuming the measurement is shot-noise limited ($Z \Glo \gg 2 \kB T$). We indeed experimentally see the background of the power spectral density increase with $\Philo$. We typically have $\Delta \omega = 2\pi (50 \, \text{Hz})$ and $\text{SNR} \approx 10^{2} \, \text{to} \, 10^{4}$. For these parameters, the Johnson-Nyquist noise in \ref{eq:noisepower} is at the $-150 \, \text{dBm}$ level, whereas our noise background is at the $-120 \, \text{dBm}$ level and the signals are at the $-80 \, \text{dBm}$ to $-110 \, \text{dBm}$ level for $24.5 \, \text{dBm}$ microwave drive power at the signal generator and a constant bias voltage of $80 \, \text{V}$. We send the optical signal to a microwave-shielded room to reduce microwave crosstalk.

\begin{figure}[ht]
\includegraphics{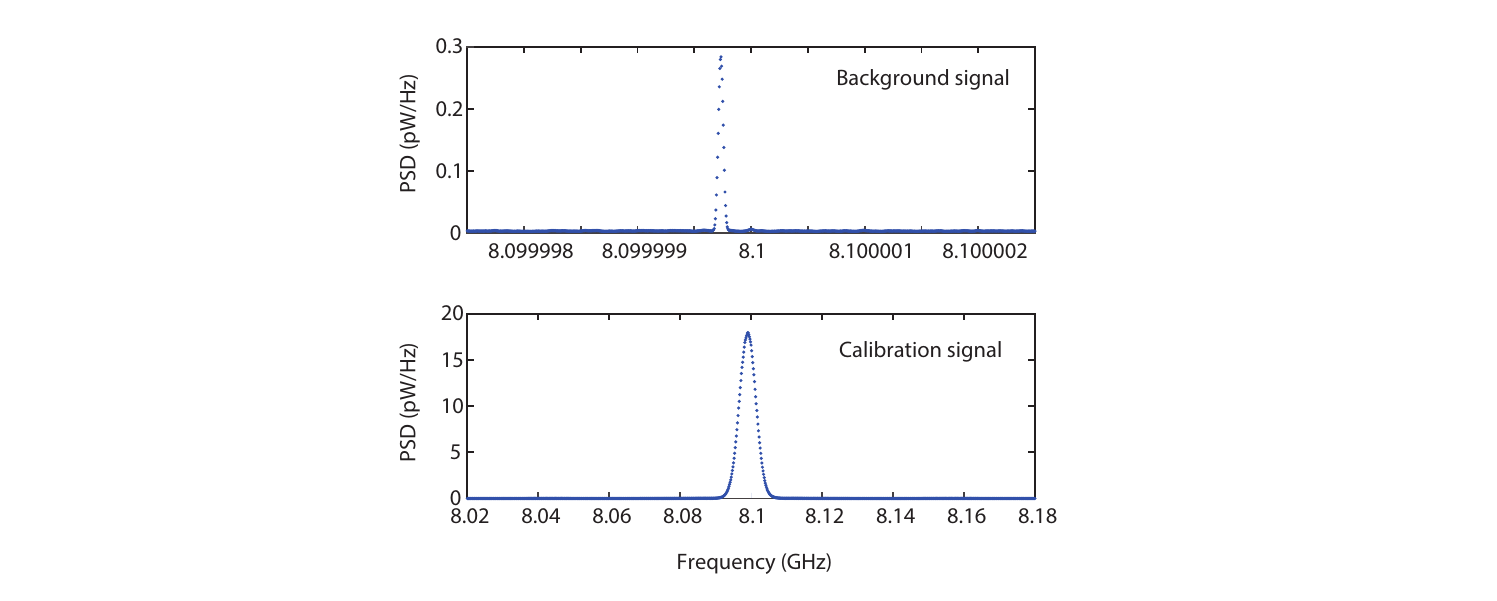}
\caption{\textbf{Calibration of optical phase fluctuations.} Top: example of an actual signal. The signal is the power spectral density of the photocurrent. The peak around 8.1 GHz originates from electrically-driven phase modulation of the optical field passing through the photonic waveguide. Bottom: example of the calibration signal. Here the peak stems from the beat note between the probe laser and a second, heavily attenuated laser with known power. The calibration is done immediately after measuring an actual signal by turning off the electrical drives. The beat note is much broader as the two lasers are not locked. We calibrate the phase fluctuations $\delta \phi$ by integrating both power spectral densities and taking their ratio as in equation \ref{eq:calibratio}. We run this calibration procedure multiple times and obtain similar results, even with several weeks in between measurements and rebuilding the measurement setup. The values in the main paper stem from a calibration at 6.5 GHz: significantly below the mechanical resonance at 7.2 GHz.}
\label{fig:calibration}
\end{figure}

\subsection{Calibration of optical phase fluctuations}
  
The signals are much stronger than the noise background so \ref{eq:noisepower} simplifies to
\begin{equation}
\label{eq:noisepower}
\int^{\Omega + \Delta \omega/2}_{\Omega - \Delta \omega/2} \frac{\text{d}\omega}{2\pi} \, Z S_{II}(\omega) = \frac{Z \Glo}{2\pi} \Phis
\end{equation}
Therefore, we calibrate our actual Kerr and electromechanical signals using a known sideband photon flux $\Phis'$ (Fig.\ref{fig:calibration}):
\begin{equation}
\label{eq:calibratio}
\frac{\int^{\Omega + \Delta \omega/2}_{\Omega - \Delta \omega/2} \frac{\text{d}\omega}{2\pi} \, S_{II}(\omega)}{\int^{\Omega + \Delta \omega'/2}_{\Omega - \Delta \omega'/2} \frac{\text{d}\omega}{2\pi} \, S'_{II}(\omega)} = \frac{\Philo \Phis}{\Philo' \Phis'}
\end{equation}
where we make $\Delta \omega$ and $\Delta \omega'$ sufficiently large to capture the full signal fluxes $\Phis$ and $\Phis'$. We realize the known sideband by injecting a second laser thato is attenuated by $20 \, \text{dB}$ red-detuned from $\omegalo$ by $\Omega$ -- using a wavemeter with picometer accuracy to set the laser wavelength. Next we measure $\Philo$ and $\Phis$ at various positions throughout the measurement setup to calibrate the optical loss induced by the chip. The beat note between the lasers is broad, requiring $\Delta \omega' = 2\pi (160 \, \text{MHz})$. We typically have $\Phis \approx 10^{8} \, \text{/s}$, $\Phis' \approx 4 \cdot 10^{14} \, \text{/s}$ and $\Philo \approx \Philo' \approx 2 \cdot 10^{16} \, \text{/s}$ at the high-speed photodetector. Next, we develop a model to predict $\Phis$. To this end, we give a derivation of the optomechanical overlap integrals.

\subsection{Optomechanical overlap integrals}

First-order perturbation theory of Maxwell's equations with respect to changes in permittivity $\dyad{\delta \epsilonr}$ shows that
\begin{equation}
\delta \neff = - \frac{\ngr}{\omega} \delta \omega
\end{equation}
with $\ngr$ the optical group index and
\begin{equation}
\delta \omega = - \frac{\omega}{2} \frac{\braket{\E|\dyad{\delta\epsilonr} |\E}}{\braket{\E|\epsilonr |\E}}
\end{equation}
with $\E[\omega]$ the unperturbed complex optical field. For our device $\ngr \approx 4.1$. We reduce the volume integrals to surface integrals
\begin{align}
\frac{\braket{\E | \dyad{\delta \epsilonr}| \E}}{\braket{\E |\epsilonr| \E}} = \frac{\intdASu \,  \E^{\star} \cdot \dyad{\delta\epsilonr} \cdot \E}{\intdAS \, \epsilonr |\E|^2}
\end{align}
as the waveguide has translational symmetry besides the periodic suspensions. Therefore we have
\begin{equation}
\delta \neff = \frac{\ngr}{2} \frac{\intdASu \, \E^{\star} \cdot \dyad{\delta \epsilonr} \cdot \E}{\intdA \, \epsilonr |\E|^2}
\end{equation}
There are three contributions to the integral in the numerator. First, the Kerr effect and photoelasticity shift the optical permittivity inside the bulk silicon and silicon dioxide via the dependence $\dyad{\delta \epsilonr}(\Emu,\um)$ on the microwave electrical field $\Emu = \Eb + \delta \E$ and the mechanical field $\um$. Second, the moving interfaces influence $\delta \neff$ via the dependence of the integration domain $S(\um)$ on the mechanical field $\um$.

Next, we derive expressions for $\dyad{\delta\epsilonr}$ resulting from each of these three mechanisms. Each of the mechanisms leads to an oscillating effective index $\delta\neff \propto \cos{(\Omega t + \varphi)}$ with some phase delay $\varphi$. Thus we have
\begin{equation}
|\delta\neff| = |\delta\neff[\Omega]|
\end{equation}
with
\begin{equation}
\delta\neff[\Omega] = \frac{\ngr}{2} \frac{\intdASu \, \E^{\star} \cdot \dyad{\delta \epsilonr}[\Omega] \cdot \E}{\intdAS \, \epsilonr |\E|^2}
\end{equation}
Here, we define $V[\Omega]$ as the Fourier component in
\begin{equation}
V(t) = \frac{V[\Omega]}{2}e^{-i\Omega t} + \text{c.c}
\end{equation}
for each variable $V$. The actual Fourier-transform $V(\omega)$ of $V(t)$ is
\begin{equation}
V(\omega) = \int_{-\infty}^{\infty} \text{d}t \, V(t)e^{i\omega t} = \pi V[\Omega]\delta(\omega - \Omega) + \pi V[-\Omega]\delta(\omega + \Omega)
\end{equation}
with $\delta(\omega) = \frac{1}{2\pi} \int_{-\infty}^{\infty}  \text{d}t \, e^{i\omega t}$ the Dirac delta distribution and $V^{\star}[\Omega] = V[-\Omega]$.

First, for the Kerr effect we have
\begin{equation}
\left.\dyad{\delta \epsilonr} [\Omega]\right|_{\text{K}} = \frac{3}{2}\dyad{\chi^{(3)}}(\omega - \Omega;\omega;\omegab;\omegamu) \cdot \Eb[\omegab] \cdot \delta \E[\omegamu]
\end{equation}
with $\Eb$ and $\delta \E$ the microwave fields generated by the voltages $\Vb$ and $\delta V$ (see section \ref{sec:electricalexcitation}). Expressed in scalar components this yields
\begin{equation}
\left.\delta \epsilonrij[\Omega]\right|_{\text{K}} = \frac{3}{2}\sum_{kl} \chi^{(3)}_{ijkl} E_{\text{b,k}}[\omegab] \delta E_{l}[\omegamu] 
\end{equation}
with $\chi^{(3)}_{ijkl}$ the third-order nonlinear susceptibility tensor.

Second, for the photoelasticity we have
\begin{equation}
\left.\dyad{\delta \epsilonr}[\Omega]\right|_{\text{p.e.}} = - \epsilonr^{2} \, \dyad{p}(\omega - \Omega;\Omega) \cdot \dyad{S}[\Omega]
\end{equation}
with $\dyad{p}$ the photoelastic tensor and $\dyad{S}$ the mechanical strain. Expressed in components this becomes
\begin{equation}
\left.\delta \epsilonrij[\Omega]\right|_{\text{p.e.}} = -\epsilonr^{2} \sum_{kl} p_{ijkl} S_{kl}[\Omega]
\end{equation}
with $S_{kl}[\Omega] = \frac{1}{2}\left(\partial_{k}u_{l}[\Omega] + \partial_{l}u_{k}[\Omega] \right)$ and $\um$ the mechanical displacement field with maximal value $\text{max}(\um) = \delta x$ (see section \ref{sec:electricalexcitation}).

Third, the moving interfaces yield a contribution
\begin{equation}
\intdASu \, \E^{\star} \cdot \dyad{\delta \epsilonr}[\Omega] \cdot \E = \epsilon_0^{-1} \int_{\mathcal{C}} \text{d}l \, u_{\text{n}}[\Omega] \, \left(\Delta \epsilon |\E_{||}|^{2} - \Delta \epsilon^{-1} |\D_{\perp}|^{2} \right)
\end{equation}
with $\mathcal{C}$ a curve capturing the interfaces, $u_{\text{n}}$ the component of the displacement field $\um$ normal to the interface and pointing towards the medium with permittivity $\epsilon_{\text{o}}$, $\Delta \epsilon = \epsilon_{\text{i}} - \epsilon_{\text{o}}$, $\Delta \epsilon^{-1} = \epsilon^{-1}_{\text{i}} - \epsilon^{-1}_{\text{o}}$ the changes in permittivity at the interfaces, $\E_{||}$ the electrical field parallel to the interface and $\D_{\perp} = \epsilon \E_{\perp}$ with $\E_{\perp}$ the electrical field perpendicular to the interface.

Finally, we define the interaction strengths as follows. The background Kerr senstivity is
\begin{equation}
\left. \partialVsq \neff \right|_{\text{K}} = \frac{\ngr}{2} \frac{\intdAS \, \frac{3}{2}\sum_{ijkl}\chi^{(3)}_{ijkl} E^{\star}_{i}E_{j} e_{b,k}[\omegab]\delta e_{l}[\omegamu]}{\intdAS \, \epsilonr |\E|^{2}}
\end{equation}
with the normalized microwave fields $\vecb{e}[\omega] = \E[\omega]/V[\omega]$. The optomechanical sensitivity is
\begin{equation}
\partialx \neff[\Omega] = \left. \partialx \neff[\Omega] \right|_{\text{p.e.}} + \left. \partialx \neff[\Omega] \right|_{\text{m.b.}}
\end{equation}
First, the photoelastic contribution is
\begin{equation}
 \left. \partialx \neff[\Omega] \right|_{\text{p.e.}} = -\frac{\ngr\epsilonr^{2}}{2} \frac{\intdAS \, \sum_{ijkl}p_{ijkl}E^{\star}_{i}E_{j} s_{kl}[\Omega]}{\intdAS \, \epsilonr |\E|^{2}}
\end{equation}
with the normalized strain
\begin{equation}
s_{kl}[\Omega] = \frac{1}{2}\left(\partial_{k}q_{l}[\Omega] + \partial_{l}q_{k}[\Omega] \right)
\end{equation}
and normalized displacement field $\vecb{q} = \um/\text{max}(\um)$. And second, the moving boundary contribution is
\begin{equation}
\left. \partialx \neff[\Omega] \right|_{\text{m.b.}} = \frac{\ngr}{2}\frac{\epsilon^{-1}_0\intdAS \, q_{\text{n}}[\Omega] \left(\Delta \epsilon |\E_{||}|^{2} - \Delta \epsilon^{-1} |\D_{\perp}|^{2} \right)}{\intdAS \, \epsilonr |\E|^{2}}
\end{equation}

\subsection{Sideband conversion efficiency}

Together, the above contributions to the overlap integrals generate optical phase fluctuations $\delta \phi$. In particular, for small phase fluctuations $\delta \phi$ we have
\begin{equation}
\alphalo e^{i\delta \phi(t)} \approx \alphalo \left(1 + i \delta \phi(t) \right)
\end{equation}
where we take $\alphalo$ to be the carrier amplitude at the output of the waveguide. The phase fluctuations $\delta\phi(t) = |\delta \phi|\cos{(\Omega t)}$
generate two sidebands onto the optical field. The fiber Bragg filter rejects one of these sidebands. Comparing with \ref{eq:opticsfluctuation} we obtain
\begin{equation}
\alphas = i \alphalo \frac{\delta \phi[\Omega]}{2}
\end{equation}
such that
\begin{equation}
\Phis = \frac{|\delta \phi|^2}{4} \Philo = \eta \Philo
\end{equation}
with
\begin{equation}
\eta = \frac{|\delta \phi[\Omega]|^2}{4}
\end{equation}
the sideband conversion efficiency. From our phase-calibration we obtain typically $\eta \approx 10^{-8}$ such that $|\delta \phi| \approx 10^{-4}$.

The phase fluctuations $\delta\phi$ stem from two sources: the broadband background Kerr effect and the narrowband optomechanical effect
\begin{equation}
\delta\phi = \delta\phi_{\text{b}} + \delta\phi_{\text{m}}
\end{equation}
The broadband background phase fluctuations are given by
\begin{equation}
\label{eq:phaseb}
\delta\phi_{\text{b}}[\Omega] = k_0 L_{\text{b}} \partialVsq \neff \frac{|\Vb|}{2} \delta V[\Omega - \omegab]
\end{equation}
whereas the narrowband mechanical phase fluctuations are
\begin{equation}
\delta\phi_{\text{m}}[\Omega] = k_0 L_{\text{s}} \partialx \neff \delta x[\Omega]
\end{equation}
The mechanical motion $\delta x[\Omega]$ is given by
\begin{equation}
\delta x[\Omega]  = \chim[\Omega] (\partialx C) \frac{|\Vb|}{2} \delta V[\Omega - \omegab]
\end{equation}
with the mechanical susceptibility
\begin{align}
\chim[\Omega] &= \frac{1}{\meff\left(\omegam^2 - \Omega^2 - i \Omega \kappam\right)} \\
&\approx \frac{\Qm}{\keff} \frac{1}{\left(\frac{\omegam^{2} - \Omega^{2}}{\kappam \Omega}\right) - i} \\
&\approx \frac{\Qm}{\keff} \frac{1}{-2\Deltar - i} = \frac{\Qm}{\keff} \mathcal{L}(\Deltar)
\end{align}
with $\Qm = \omegam/\kappam$ the mechanical quality factor and the Lorentzian
\begin{equation}
\mathcal{L}(\Deltar) =  \frac{1}{-2\Deltar - i}
\end{equation}
Here, we define the relative detuning from the mechanical resonance $\Deltar = (\Omega^{2} - \omegam^{2})/(2\kappam \Omega) \approx (\Omega - \omegam)/\kappam$. The latter approximation holds close to the mechanical resonance. Thus the sideband conversion efficiency can be written as
\begin{align}
\eta &= \frac{|\delta\phi_{\text{b}}|^2}{4} \left|1 + \frac{\delta\phi_{\text{m}}}{\delta\phi_{\text{b}}}\right|^{2} \\
&= \etab |1 + r \mathcal{L}(\Deltar) |^{2}
\end{align}
with $\etab = |\delta\phi_{\text{b}}|^{2}/4$ the background conversion efficiency and $r$ a dimensionless ratio that captures the relative strengths of the non-resonant background and the resonant mechanical effect
\begin{equation}
r = \frac{\Qm}{\keff} \frac{(\partialx C)(\partialx \neff)}{\partialVsq \neff} \frac{L_{\text{s}}}{L_{\text{b}}}
\end{equation} 
The shape of this resonance is identical to the Fano curves measured in optically-driven cross-phase modulation and wavelength-conversion. Its properties are discussed in the appendix of \cite{VanLaer2015b}. Besides the three contributions to $\delta \neff$ derived in the above, other mechanisms may contribute to the sideband conversion as well \cite{Schriever2015}. These include symmetry breaking by surfaces or strain, the bulk quadrupolar $\chi^{(2)}$ as well as free-carrier phase and amplitude modulation. The first two mechanisms have a broadband response, while the latter two mechanisms have a strong dependence on the modulation frequency $\Omega$. These mechanisms interfere with different phases. Thus we perform our fits to
\begin{equation}
\label{eq:fano}
\eta = \etab|1 + re^{i\varphi}\mathcal{L}(\Deltar)|^{2}
\end{equation}
with $\varphi$ an additional phase of the mechanically-mediated phase-modulation with respect to the background.

The above derivations assume sum-frequency driving (SFD) with two separate fluctuating bias and drive voltages $\Vb(t)$ and $\delta V(t)$. However, the electrically-induced phase fluctuations are a factor 2 stronger in the case of a constant bias voltage $\Vb$. In particular, the above curve shape of equation \ref{eq:fano} remains identical but $\etab$ is a factor 4 larger since 
\begin{equation}
\delta \phi_{\text{b}}[\Omega] = k_0 L_{\text{b}} \partialVsq \neff \Vb \delta V[\Omega]
\end{equation}
replaces equation \ref{eq:phaseb} when $\omegab = 0$. The other parameters are not affected. Similarly, the efficiency drops by a factor 4 with respect to SFD in the case of degenerate second-harmonic driving (SHD) with only one fluctuation voltage $\delta V$.

\subsection{Properties of $\dyad{\chi}^{(3)}$ and $\dyad{p}$}

Here, we discuss a few important properties of the Kerr tensor $\chi^{(3)}_{ijkl}$ and the photoelastic tensor $p_{ijkl}$. Silicon has the diamond cubic structure with point group $m\overline{3}m$. Thus $\chi^{(3)}_{ijkl}$ has 21 non-zero elements of which 4 are independent. However, since we are interested in frequencies far away from the bandgap the Kleinman symmetry applies so there are only two independent parameters: $\chi^{(3)}_{xxxx}$ and $\chi^{(3)}_{xxyy}$. For a range of frequencies below the bandgap of silicon it was found that $\chi^{(3)}_{xxyy,\text{Si}} \approx \frac{1}{2.36} \chi^{(3)}_{xxxx,\text{Si}}$. Thus only $\chi^{(3)}_{ijkl}$ with $ijkl$ equal to $iiii$ or a permutation of $iijj$ are non-zero and all values follow from $\chi^{(3)}_{xxxx}$. Measurements found $\chi^{(3)}_{xxxx,\text{Si}} =  2.45 \cdot 10^{-19} \, \text{m}^{2}/\text{V}^{2}$ at $1550 \, \text{nm}$ \cite{Osgood2009,Timurdogan2017}. There is some dispersion in this value around the two-photon bandgap $2200 \, \text{nm}$, but we expect $\chi^{(3)}_{xxxx}$ to be similar at the microwave frequencies involved in this work. This value is consistent with the equation
\begin{equation}
n_{2,\text{Si}} = \frac{3}{4\epsilon_0 c n^{2}_{\text{Si}}} \chi^{(3)}_{\text{eff,Si}}
\end{equation}
given $n_{2,\text{Si}} = 5 \cdot 10^{-18} \, \text{m}^{2}/\text{W}$. Similarly, thermal oxide is isotropic such that $\chi^{(3)}_{xxyy,\text{SiO}_2} = \frac{1}{3} \chi^{(3)}_{xxxx,\text{SiO}_2}$ with $\chi^{(3)}_{xxxx,\text{SiO}_2} =  2.5 \cdot 10^{-22} \, \text{m}^{2}/\text{V}^{2}$ and $n_{2,\text{SiO}_2} = 3.3 \cdot 10^{-20} \, \text{m}^{2}/\text{W}$. Thus we have
\begin{align}
&\sum_{ijkl} \chi^{(3)}_{ijkl} E^{\star}_{i}E_{j} e_{b,k}[\omegab]\delta e_{l}[\omegamu] = \\
& \left(\chi^{(3)}_{xxxx} |E_{x}|^{2} + \chi^{(3)}_{xxyy}\left(|E_{y}|^{2} + |E_{z}|^{2}\right)\right) e_{b,x}[\omegab]\delta e_{x} [\omegamu]+\notag \\
& \left(\chi^{(3)}_{xxyy} \left( |E_{x}|^{2} + |E_{z}|^{2}\right) + \chi^{(3)}_{xxxx} |E_{y}|^{2}\right) e_{b,y}[\omegab]\delta e_{y}[\omegamu]+\notag \\
& 2\chi^{(3)}_{xxyy} \Re{(E_{x}E^{\star}_{y})}  \left(e_{b,x}[\omegab]\delta e_{y}[\omegamu] + e_{b,y}[\omegab]\delta e_{x}[\omegamu]\right) \notag
\end{align}
Similarly, the photoelastic tensor of silicon has three independent components $p_{1111} \equiv p_{11} = -0.09$, $p_{1122} \equiv p_{12} = 0.017$ and $p_{1212} \equiv p_{44} = -0.05$ in contracted notation. Therefore we have
\begin{align}
&\sum_{ijkl}p_{ijkl}E^{\star}_{i}E_{j} s_{kl}[\Omega] = p_{11}|E_{x}|^{2} s_{xx}[\Omega] \\
& p_{11} |E_{y}|^{2} s_{yy}[\Omega] + p_{12} |E_{y}|^{2} s_{xx}[\Omega] \notag \\
& + p_{12} |E_{x}|^{2} s_{yy}[\Omega] + 2p_{44}\Re{(E^{\star}_{x}E_{y})} s_{xy}[\Omega] \notag
\end{align}
where we used $s_{xz} = s_{yz} = s_{zz} = 0$ for our $\Gamma$-point mechanical mode.

\begin{figure}[ht]
\includegraphics{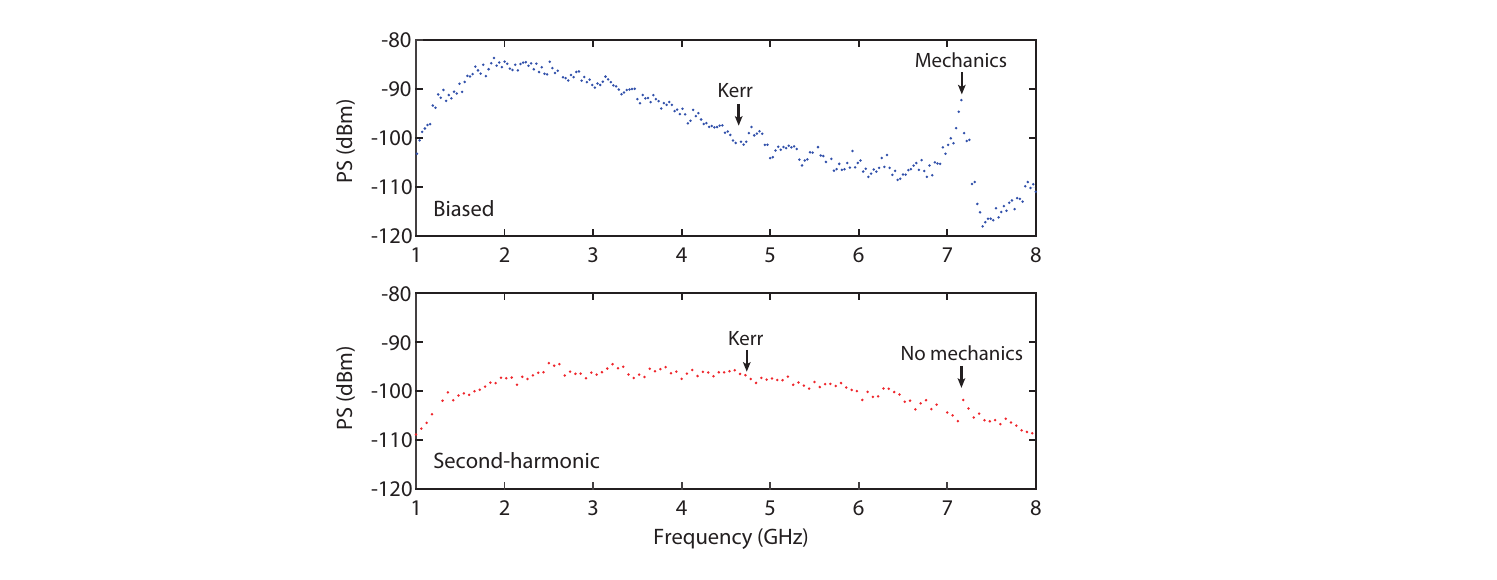}
\caption{\textbf{Biased vs. second-harmonic driving of the optical phase fluctuations.} Top: measured power spectrum of the photocurrent when driving electrically with a constant bias field and a microwave field at the displayed frequency. Bottom: measured power spectrum of the photocurrent when driving electrically without a constant bias field but with a microwave field at half the displayed frequency. We do not observe the mechanically-mediated phase fluctuations in the second-harmonic signal. In addition, the background effect is of roughly similar magnitude in the bottom figure even though the oscillating voltage is significantly weaker than the bias voltage.}
\label{fig:shdriving}
\end{figure}

\section{Sum-frequency driving of the optical phase fluctuations}
\label{sec:SFD}

The measurements presented in the main paper are made with a constant bias field and an oscillating microwave field. However, as derived in appendix \ref{sec:electricalexcitation} the driving forces scales as $V^{2}$ so it is possible to drive the background Kerr effect and the mechanical oscillator with two oscillating voltages at $\omega$ and $\omegab$ if their frequencies are chosen appropriately such that $\omega + \omegab = \Omega$.  We call this type of actuation sum-frequency driving (SFD) when both frequencies fall below the driving frequency $\Omega$. This enables us to inject two electrical driving fields with frequencies $\omega$ and $\omegab$ both above the RC-cutoff $\omega_{\text{RC}}/(2\pi) \approx 51\,\text{MHz}$ of the silicon waveguide (see section \ref{sec:devphys}). Thus both microwave fields can penetrate the bulk silicon. We call this approach second-harmonic driving (SHD) when the two fields are identical and $2\omega = 2\omegab = \Omega$.

We perform a series of SHD and SFD measurements in absence of a strong bias field. None of these measurements exhibit a clear mechanical signal (Fig.\ref{fig:shdriving}). All measurements indicate the presence of a strong Kerr background. In addition, the background power spectrum is of roughly similar magnitude as in the biased measurements. This indicates that the background Kerr parameter $\partialVsq \neff$ seen in the SFD and SHD traces is about a factor $4\Vb/\delta V \approx 4 \times 80/5.3 \approx 60$ stronger than in the biased measurements. This is consistent with the absence of mechanical signals in the SFD/SHD measurements (Fig.\ref{fig:shdriving}) assuming the electromechanical coupling strength $\partialx C$ does not increase significantly.

\end{document}